\documentstyle{article}

\def\ben{\begin{equation}}
\def\een{\end{equation}}
\def\bena{\begin{eqnarray}}
\def\eena{\end{eqnarray}}

\input amssym.def
\input amssym.tex
\begin{document}

\hfuzz=100pt
\title{Two loop and all loop finite 4-metrics} 
\author{G W Gibbons 
\\
D.A.M.T.P.,
\\ Cambridge University, 
\\ Silver Street,
\\ Cambridge CB3 9EW,
 \\ U.K.}
\maketitle

\begin{abstract} In pure Einstein theory,
Ricci flat Lorentzian 4-metrics of Petrov types III or N
have vanishing counter terms up to and including two loops.
Moreover for pp-waves and type-N spacetimes of Kundt's class
which admit a non-twisting, non
expanding, null congruence all possible invariants
formed from the Weyl tensor and its covariant derivatives  vanish.
Thus these Lorentzian metrics suffer no quantum corrections 
to all loop orders.
By contrast 
for complete non-singular  Riemannian metrics the two loop counter term vanishes only if the metric is flat.

\end{abstract}

Solutions of classical field equations for
which the counter terms required to regularize 
quantum fluctuations 
vanish are of particular importance because
they offer insights into the behaviour of the full quantum theory.
Conversely, classical solutions
for which quantum corrections do not vanish
afford little or no insight
into the strongly quantum regime since quantum corrections 
are expected to be large.
Thus in gravity theories, in plane wave spacetimes, which are of Petrov type N, all counter terms vanish on shell and so they 
suffer no quantum corrections \cite{GG,D}, This corresponds 
to  the fact that the graviton remains massless
in the quantum theory and presumably remains a physically
valid concept no matter how large quantum effects are. 

Another important class
of examples occur in supersymmetric theories
when one has a supersymmetric, so-called BPS,
solution of the classical equations. In four-dimensional
supergravity theories, examples of BPS solutions are provided by
positive definite metrics whose curvature is self-dual or anti-self dual,
or equivalently $SU(2) \equiv Sp(1)$ holonomy.
An interesting question is whether there are non-trivial
positive definite metrics
which are not supersymmetric and for which quantum corrections
nevertheless vanish in pure gravity. One purpose of the present note
is to show that, subject to regularity, there are none.

The coefficients of quantum corrections
to Ricci flat solutions
of Einstein's theory of gravity in four dimensions
have been calculated up to two loops. However it  is widely believed
that there are non-vanishing terms to arbitrarily high
loop order.
The one loop term
counter term is proportional to  quadratic invariant
of the curvature tensor $I_2= R_{abcd}R^{abcd}$
the integrand of the Gauss-Bonnet theorem. 
If the metric is Lorentzian and of Petrov types
III or N, then this invariant vanishes \cite{Z}.
However, if the   background
has a positive definite metric, then the counter term $I_2$
is positive definite
and can  only vanish
if the background is flat. 
However because it is a total derivative, one might think that
this fact may not be important, and could  perhaps be ignored. 
Even if this point is conceded one still has two consider higher looops.
Let us therefore turn to the two loop term
which is proportional to the cubic invariant of the curvature tensor
$I_3=R_{ab}\thinspace ^{cd} R_{cd} \thinspace ^{ef} R_{ef} \thinspace ^{ab}$
\cite{GS,V}. If the background is Lorentzian and of Petrov types III
or N then this invariant also vanishes \cite{Z}. 
Indeed in a Ricci flat spacetime of Petrov types III or N
all invariants formed solely by contracting products
of the  Riemann tensor using the metric must necessarily vanish.
This property, which holds only for these Petrov types, is clear
from the expression for the Weyl Spinor $C_{ABCD}$ in
terms of the principle spinors $\kappa ^A$ and $\iota^A$.
For type N
\ben C_{ABCD}= \iota_A \iota_B\iota_C\iota_D. 
\een
and  for type III
\ben C_{ABCD}= \iota_A \iota_B \iota_{(C} \kappa_{D)},
\een
and no non-vanishing contractions are possible.

Even if the background metric is positive definite,
a  cubic invariant cannot be  positive definite 
and so it makes sense to ask:
are there
non-singular backgrounds $\{M, g\}$
for which it vanishes? Note that there is no analogue
of Petrov type III or N for positive definite metrics \cite{K}. However
it is possible to give a similar algebraic characterization
of the vanishing of the cubic invariant.
The self-dual and anti-self-dual parts of the Weyl tensor 
determines two symmetric trace-free symmetric matrices
$D^\pm_{ij}$ say. The invariant $I_3$ is proportional to 
\ben
{\rm tr} (D^+)^3 + {\rm tr} (D^-)^3. 
\een
A short calculation reveals that $I_3$ will vanish
if and only if both  $D^+$ and $D^-$ have a zero eigen-value.

Remarkably, the answer to the question asked above
is provided, almost word for word,
by a rather old  result of Lichnerowicz \cite{L1,L2} 
(see also \cite{TV, MB}). 
Lichnerowicz derives an identity,  valid in all dimensions
if the Ricci tensor vanishes, which 
reads
\ben
\nabla ^2 I_2= 2\nabla _e R_{abcd} \nabla ^e R^{abcd} +K. \label{ident}
\een
where the scalar (as opposed to pseudo-scalar)
 $K$ is a cubic invariant of the curvature
tensor.  In four dimensions $K$ must obviously be a multiple of $I_3$
since if the Ricci 
tensor vanishes that is the only non-vanishing scalar cubic
invariant. 

One may now integrate (\ref{ident}) over the manifold $M$. 
The left hand side gives a boundary term, which obviously vanishes if 
$M$ is closed and which one may check will certainly
vanish if $M$ is ALE
or ALF. We deduce that if $I_3=0$ then
\ben
\nabla_e R_{abcd}=0 \label{symmetric}.
\een
In fact (\ref{symmetric}) says that $M$ is 
locally at least a symmetric space. 
However it is clear that one may deduce more. Since 
\ben
\nabla_e I_2=0, 
\een
it follows in the ALF or ALE case that $I_2=0$ everywhere, since it 
vanishes at infinity. But if $I_2=0$ the metric must be locally flat.
If $M$ is closed, then we need a more complicated argument.
However 
the conclusion is the same: the metric must be locally flat \cite{L1}.

Thus even
self-dual metrics,  are not exact and would  receive quantum corrections
in pure Einstein theory at two loops.
By contrast, non-trivial vacuum Lorentzian metrics of Petrov type III  and N, which are not  plane waves or pp-waves certainly exist \cite{EX}.
In the case of type N vacuum spacetimes, it is known \cite{BJ}
that provided they admit a non-expanding, non-twisting geodesic null congruence, then all invariants formed from the Weyl tensor
and it's covariant derivatives vanish. It seems that this class of solutions,
which belong to Kundt's class \cite{EX}, do not correspond to
our conventional idea of a graviton since gravitons are usually taken to be
described by pp-waves. In view of the vanishing of all
quantum corrections it would seem worthwhile investigating these
metrics further. Since pp-waves admit a covariantly constant
spinor, they have a holonomy group which is not the full
Lorentz group. An interesting question is whether
the other all loop finite metrics in Kundt's class also have a
reduced holonomy group.

If one drops the non-expanding condition but retains the
non-twisting condition then there are non-vanishing invariants.
For example, in \cite{BJ} the invariant
\ben
C^{abcd;ef}C_{amcn;ef}C^{lmrn;st}C_{lbrd;st}
\een
is shown not to vanish. The situation with respect to type III spacetimes
is not known to the author.

We close with the remark that it would be
interesting to extend these results
 when a cosmological term is present, or to higher dimensions.
In the latter case Lichnerowicz's identity holds in all dimensions.
However the structure of the counter terms and possible invariants
 will be different.

\end{document}